\documentclass{nature}
\usepackage{color} % FIXME: remove
\usepackage{graphicx} % FIXME: remove
\usepackage[usenames,dvipsnames]{xcolor} % FIXME: remove

\title{Contact activity and dynamics of the online elite}

\author{Enys Mones$^1$, Arkadiusz Stopczynski$^{1,2}$ \& Sune Lehmann$^{1,3}$}

\begin{document}

% FIXME: remove after finalization
\newcommand{\enys}[1]{\textcolor{blue}{\textsc{Enys}: #1}}
\newcommand{\arek}[1]{\textcolor{magenta}{\textsc{Arek}: #1}}
\newcommand{\sune}[1]{\textcolor{green}{\textsc{Sune}: #1}}
\newcommand{\todo}[1]{\textcolor{red}{#1}}
\newcommand{\goal}[1]{\textcolor{orange}{\textsc{goal}: #1}}
\newcommand{\highlight}[1]{\textcolor{purple}{#1}}

\maketitle

\begin{affiliations}
 \item Department of Applied Mathematics and Computer Science, Technical University of Denmark, Kgs. Lyngby, Denmark
 \item Media Lab, Massachusetts Institute of Technology, Cambridge, MA, USA
 \item The Niels Bohr Institute, University of Copenhagen, Copenhagen, Denmark
\end{affiliations}

\begin{abstract}
Humans interact through numerous channels to build and maintain social connections: they meet face-to-face, initiate phone calls or send text messages, and interact via social media.
Although it is known that the network of physical contacts, for example, is distinct from the network arising from communication events via phone calls and instant messages, the extent to which these networks differ is not clear.
In fact, the network structure of these channels shows large structural variations.
Each network of interactions, however, contains both central and peripheral individuals: central members are characterized by higher connectivity and can reach a high fraction of the network within a low number of connections, contrary to the nodes on the periphery.
Here we show that the various channels account for diverse relationships between pairs of individuals and the corresponding interaction patterns across channels differ to an extent that hinders the simple reduction of social ties to a single layer.
Furthemore, the origin and purpose of each network also determine the role of their respective central members: highly connected individuals in the person-to-person networks interact with their environment in a regular manner, while members central in the social communication networks display irregular behavior with respect to their physical contacts and are more active through rare, social events.
These results suggest that due to the inherently different functions of communication channels, each one favors different social behaviors and different strategies for interacting with the environment.
Our findings can facilitate the understanding of the varying roles and impact individuals have on the population, which can further shed light on the prediction and prevention of epidemic outbreaks, or information propagation.

\end{abstract}

% introduction
\section*{Introduction}
In modern society, more and more modes of communication are available, often covering different aspects of our lives: we meet others face-to-face to build and maintain social ties\cite{eubank2004modelling,cattuto2010dynamics,salathe2010high,starnini2013immunization}; we make phone calls for various reasons\cite{onnela2007pnas,aledavood2015digital} (as a replacement for physical contacts or simply arranging future appointments); we interact with others on social media\cite{mislove2007measurement,ellison2007benefits,grabowicz2012socials}.
Each channel can require a different level of time commitment as well as physical effort to participate, and may correspond to social ties of different strength\cite{eagle2009pnas,sekara2014thestrength,gilbert2009predicting}.
Understanding the function of and interplay between these channels has been the subject of increased research interest over the past few years\cite{eagle2006reality,lazer2009computational,raento2009smartphones,stehle2011high}, along with a growing number of studies focusing on the understanding and quantitative analysis of multilayer networks\cite{dedomenico2013mathematical,dedomenico2015,nicosia2015measuring}. 
On one hand, the question revolves around how these channels interact and how the concurrent application of them actually affects our communication and the dynamics of our social environment.
On the other hand, as ever higher fraction of communication takes place via the digital channels, digital traces can provide unprecedented accuracy about human behavior and social interactions\cite{eagle2009pnas,aharony2011social,isella2011close,miller2012smartphone,staiano2012friends}.

A central question in the analysis of social networks is to identify the central individuals in a community solely based on their position in the global structure of the interactions\cite{freeman1978centrality,garciaherranz2014}.
While it has been shown that there are differences in how people position themselves with respect to the digital networks\cite{christakis2010social,manriquee2016women}, it remains unclear whether these differences materialize in any aspect of their physical contacts.
This raises the question: do central members of a social network have specific behavioral patterns in their physical proximity networks as well?

Here we analyze the interplay between digital networks and real-world physical contacts by analyzing the multi-channel data of more than 500 university students.
First we show that the frequency of interaction on social networks and by phone calls is not trivially correlated with the physical contacts, indicating a fundamental difference between these networks.
Furthemore, depending on the physical distance of the proximity contacts (which is related to the strength of the social tie between the actors\cite{sekara2014thestrength}), communication networks show varying levels of structural similarity with proximity networks.
As a result, we point out that the physical distance in proximity interactions provides information about the nature of the contacts, as short-range interactions resemble the communication networks more closely\cite{sekara2014thestrength}.
Finally, we quantify the fundamental differences in the behavior of central individuals based on the communication and proxmity networks.
By measuring the intensity and regularity of physical engagement with the population, we show that students central in the digital communication networks exhibit high relative activity during evenings and in the weekend, and are less predictable compared to the population average.

% results
\section*{Results}

\subsection*{Strength of ties}
Various channels of interactions can represent fundamentally different aspects of a relationship and correspond to different strength of social ties.
Phone calls and text messages are known to occur primarily between family members and acquaintances, with high call duration and frequency indicating a strong relationship\cite{sekara2014thestrength}, and mainly appear between pairs of individuals.
On the contrary, social network sites, such as Facebook or Google+, serve as a platform to maintain a wide range of social interactions from instant messaging to posts, or quick responses to events in the ego-network of the individual.
Due to the absence of substantial effort and time commitment to engage, these channels constitute a weaker form of direct communication and may suggest a weaker social link.
Here we consider the functional network of Facebook, that is, each time a student interacts with any other (via posting on wall, tagging, commenting, etc), a link of activity is formed, irrespective of the interaction type.
This is in contrast to the static web of Facebook friendship status, which does not involve active participation once the relationship is established.
Finally, physical proximity plays an essential part in maintaining relationships, being reported as the strongest impact on emotional connection as well as providing the strongest ties of high quality\cite{sherman2013theeffects,antheunis2012thequality,marsden1984measuring,mesch2006thequality}.
Nevertheless, the mere presence of proximity between individuals does not imply a social interaction (as proximity can occur without the active interaction between participants), and thus this channel cannot trivially be used for inferring social connections.
To emphasize the importance of physical distance, here we make the distinction between two types of proximity interactions\cite{stopczynski2015physical}: \emph{ambient}, corresponding to a physical distance of up to 10-15m between the participants; and \emph{intimate} that requires a distance of 1m or less (see Methods for the details on the construction of these interactions).

Altough it has been shown that social ties can be inferred from online activity (or vice versa)\cite{scellato2011exploiting,jones2013inferring,sapiezynski2016inferring,sapiezynski2016offline}, the interplay between these channels is inherently complex and it should be noted that estimating the strength of social ties by reducing them to one of the channels (or aggregating the respective networks) may have non-trivial implications.
Therefore, in this paper we focus on the understanding how observing various levels of engagement across channels can be utilized to gain insight into certain behavioral patterns.
After the overview of the structural differences and underlying correlations found in these networks and, we focus on a particular case where proximity contacts play a crucial role, namely on spreading phenomena taking place on the physical network.

\begin{figure}
\centering
\includegraphics[width=0.5\textwidth]{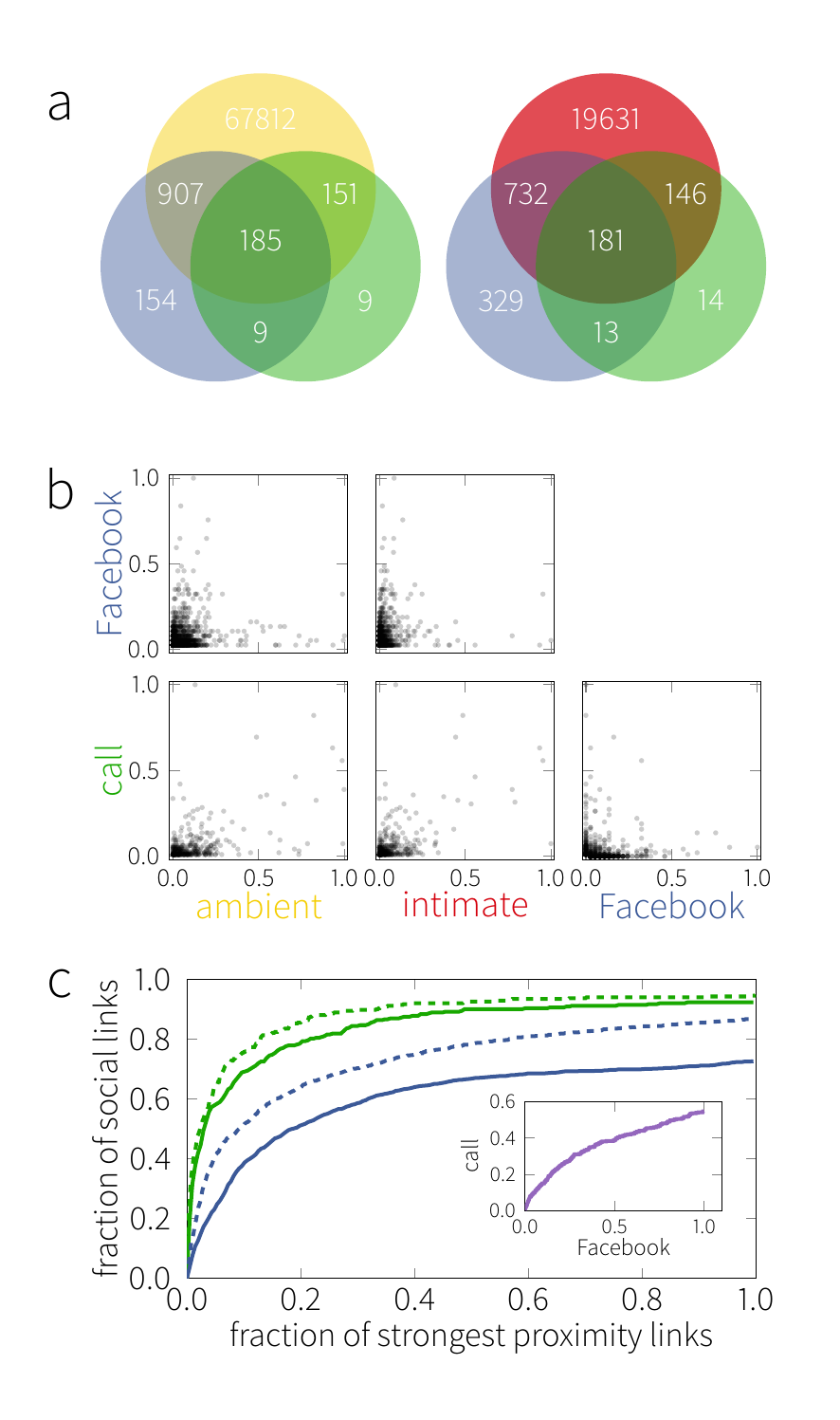}\\
\caption{\textbf{Comparison of link sets in the three networks.}
a) Venn diagrams showing the number of contacts in the different channels: ambient (left), intimate (right).
b) Correlation of tie strength between the channels.
Strength is defined by the appearance frequency of the links, normalized by the frequency of strongest link.
c) Fraction of social links recovered by the strongest ambient (dashed) and intimate (solid) physical proximity contacts: calls (green) and Facebook (blue).
Inset shows the coverage of call links by Facebook interactions.
\label{fig:networks_comparison}}
\end{figure}

Figure~\ref{fig:networks_comparison} summarizes the usage activity across channels by investigating how ties are expressed across the networks.
As we can see in Fig.~\ref{fig:networks_comparison}a, there are remarkable variations in how the various channels capture contacts between pairs of individuals.
The proximity networks contain a vast majority of the contacts (only 172 and 356 of all recorded contacts are not present in the ambient and intimate networks, respectively) and a dominant fraction of all links are exclusively represented as physical contacts (67\,812 and 19\,631, accounting for 48\% and 14\% of all possible links).
% total:	69227		21046
% FB:		1255		1255
% call:		354			354
Compared to the links observed in the proximity networks, a moderate number of interactions (1.81\% and 5.96\% of all interactions considering ambient or intimate network, respectively) are covered by Facebook activity and a negligible fraction of contacts (0.51\% and 1.68\%) are present in the phone call network as well.
The presence of call-only relationships is, in part, due to the fact that we use a one month time window and by increasing the period of observation, those interactions diminish.
Also note that the number of links present in all three channels remains around 180, irrespective the proximity channel considered.
The fact that these sets of links consist of the same pairs of users, suggests that even though the structure of the ambient network is blurred by spurious encounters, after removing those links that are not present in other channels, it is still possible to recover the strong links represented in the intimate network.

% Figure 1

While it has been indicated that intensity of digital communication not necessarily imply the strength of the corresponding social links\cite{wiese2015call}, we expect ties expressed in the phone network or Facebook interaction network to correspond to real social ties and therefore we expect these ties to be stronger (i.e., active with high frequency) in the physical proximity network\cite{sekara2014thestrength,stopczynski2015physical}.
Surprisingly, this is only partly true, as shown in Fig.~\ref{fig:networks_comparison}b, where weight of the links (number of occurrence) is plotted.
On one hand, the absence of structure in the plot of Facebook and call weights indicates that these two communication channels are used interchangeably (with a Pearson correlation of $r_\mathrm{Facebook,\,call} = 0.007$).
On the other hand, the communication networks show moderate positive correlation with the physical links: $r_\mathrm{Facebook,\,ambient} = 0.130$, $r_\mathrm{Facebook,\,intimate} = 0.146$ and $r_\mathrm{call,\,ambient} = 0.510$, $r_\mathrm{call,\,intimate} = 0.554$.
In general, call network shows higher correlation with respect to link weight with the proximity networks compared to Facebook activity, and the intimate network seems to be better predictor of strong social ties having a consistently higher correlation with communication channels than that of the ambient network.
The highest correlation is found between calls and intimate interactions.
% r(F,C) = 0.00730638
% r(F,A) = 0.129806
% r(F,D) = 0.146456
% r(C,A) = 0.510248
% r(C,D) = 0.553994

Assuming that calls and Facebook activity (i.e., \emph{online communication}) correspond to strong social ties, the capability of proximity links to predict those ties can be assessed by calculating the fraction of links in the former two networks that are covered by the strongest physical contacts, which is shown in Fig.~\ref{fig:networks_comparison}c.
The most striking observation is that once around 5\,000 of the strongest proximity links are considered, the contribution of additional links is comparably small (even negligible in case of call contacts).
That is, almost none of the remaining links correspond to links found in the Facebook or call networks.
Again, we see that call contacts can be captured more efficiently by proximity links than Facebook interactions.
However, even in the case of the intimate network, the strongest 1\,000 links cover only 58\% of all call contacts, meaning that although a large fraction of digital communication links are also included in the proximity networks (see Fig.~\ref{fig:networks_comparison}a), these links are not the necessarily the strongest physical links, and the ordered set of social ties are separated by many strong proximity links that are not represented by phone calls.
In other words, many high-frequency physical links correspond to passive and socially less significant interactions.
As the inset of Fig.~\ref{fig:networks_comparison}c illustrates, the strongest call contacts are distinct from the strongest Facebook interactions, indicating that the links characterized by the most intense communication on Facebook constitute separate group from that of the most frequent mobile calls.
In other words, individuals tend to avoid mixing the two channels and limit the maintenance of relationships to one of them.

Here we acknowledge the fact that a majority of the proximity links are due to the co-location of students attending the same classes, which explains the low correlations seen in the data.
It should be noted, however, that due to the nature of ambient and intimate networks, the latter exhibits nevertheless higher agreement with the call and Facebook networks.
In the next section, we further elaborate on this observation and show that the intimate network also shows higher level of structural similarity with the communication networks.

\subsection*{Structural similarity}
\begin{figure}
\centering
\mbox{
\includegraphics[width=1\textwidth]{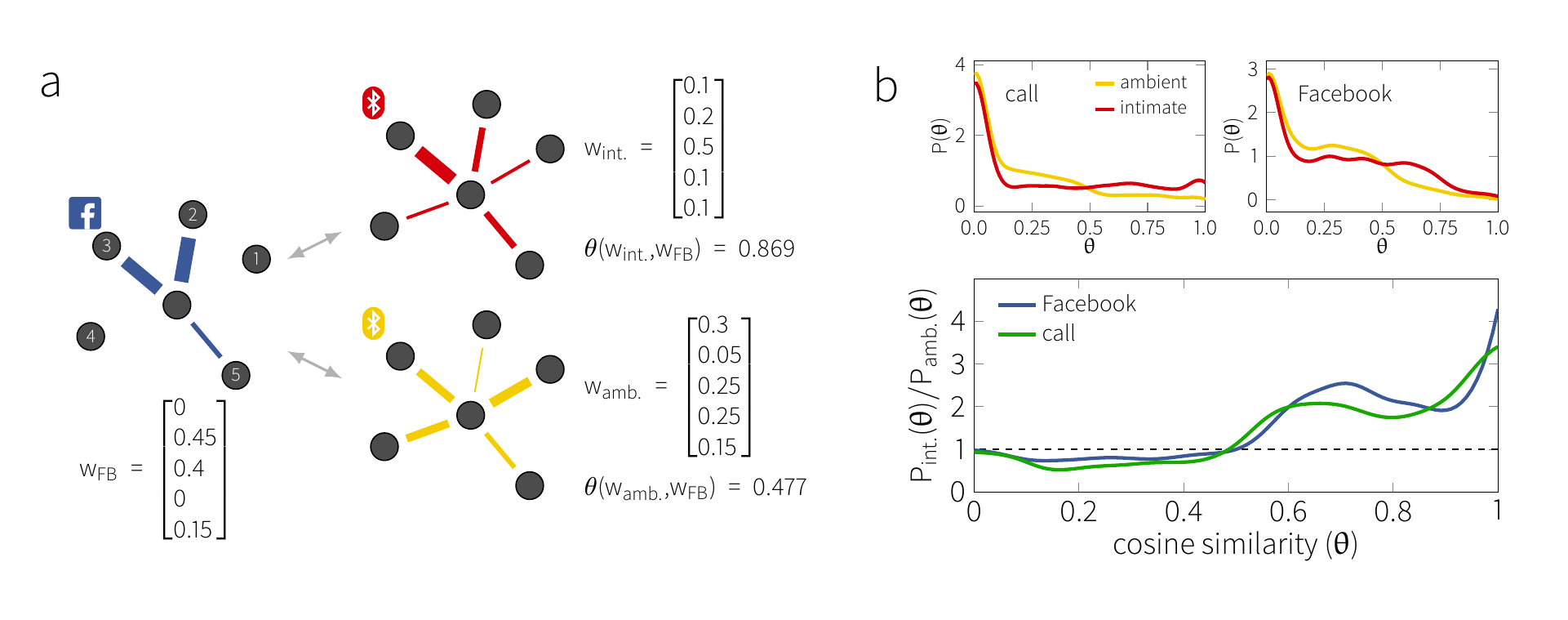}
}\\
\caption{\textbf{Structural similarity of the physical proximity and communication networks.}
a) Illustration of neighbor cosine similarity between networks.
b) Distribution of cosine similarity between the intimate network and communication networks.
Proximity networks are compared to calls (left) and Facebook interactions (right).
On the bottom, the pointwise ratio of the intimate and ambient distributions are reported.
Data is binned and a kernel-density smoothing is applied.
\label{fig:structure_comparison}}
\end{figure}

Besides single links, we can compare the local structure of the networks, that is, the ego-networks.
To this end, we calculate the similarity between the contact lists of a pair of individuals, as shown in Fig.~\ref{fig:structure_comparison}.
For a given participant $u$, we first consider their generalized neighbor-set, which consists of all other participants, and construct the weighted degree vector $w^\mathrm{u}_\mathrm{c}$ that corresponds to the distribution of interactions with $u$'s alters in channel $c$.
In other words, the weighted degree describes how the user distributes their time over their contacts in a given channel (Fig.~\ref{fig:structure_comparison}a).
Similarity between the weighted degree of a specific individual in two different channels $c$ and $c'$ is calculated using the cosine similarity:
\[
\theta(w^\mathrm{u}_\mathrm{c}, w^\mathrm{u}_\mathrm{c'}) = \frac{w^\mathrm{u}_\mathrm{c} \cdot w^\mathrm{u}_\mathrm{c'}}{\Vert w^\mathrm{u}_\mathrm{c}\Vert\Vert w^\mathrm{u}_\mathrm{c'}\Vert},
\]
where $x\cdot y$ denotes the scalar-product of vectors $x$ and $y$, while $\Vert x\Vert$ is the $\ell_2$ norm of a vector $x$.
When compared to the communication networks, the distributions $P_\mathrm{int.}(\theta)$ and $P_\mathrm{amb.}(\theta)$ characterize how similar the intimate and ambient networks are to the call and Facebook activity networks.
To quantify how different each of the proximity networks are from the communication networks, we consider the distribution of cosine similarity values, shown in Fig.~\ref{fig:structure_comparison}b.
In the top two plots we report the distribution of the similarity between the proximity networks and the digital communication networks.
For both calls and Facebook interactions, the intimate network displays higher probability density at high similarity values ($\theta > 0.5$) than the ambient network.
This observation is further emphasized in the pointwise ratio of the distributions in Fig.~\ref{fig:structure_comparison}b bottom plot.
For low values of $\theta$ ($\theta < 0.5$), there is little relative differene in the intimate and ambient networks, but in the case of high similarity ($\theta > 0.5$), the intimate network exhibits consistently higher similarity with the digital communication channels, indicating a stronger correspondence between physical and digital communication contacts.
Due to its higher similarity with the communication networks with respect to link strength and structure, in the rest of the paper we will focus on the intimate network to investigate the role of active individuals in the various channels.

% Figure 2

\subsection*{Contact patterns}
In each network, a small set of individuals can be considered as central members of their respective communities.
Proximiy networks are characterized by high link density and therefore measures based on geodesic distance fail to distinguish among individuals.
In other words, due to the high number of links, the distribution of closeness centrality or $k$-coreness is narrow, and the value of the centrality measure is not descriptive of the individual's status in the network.
Therefore, in the physical proximity network we rank students according to their total time spent in the proximity with others and choose the ones with the highest time as central.
We argue that connectivity is a meaningful measure of centrality.
In the context of epidemic monitoring, Smieszek and Salath\'{e} have shown, that the weighted degree is able to locate potential candidates for the monitoring problem with efficiency close to the optimal solution\cite{smieszek2013lowcost}.

\begin{figure}
\centering
\includegraphics[width=0.5\textwidth]{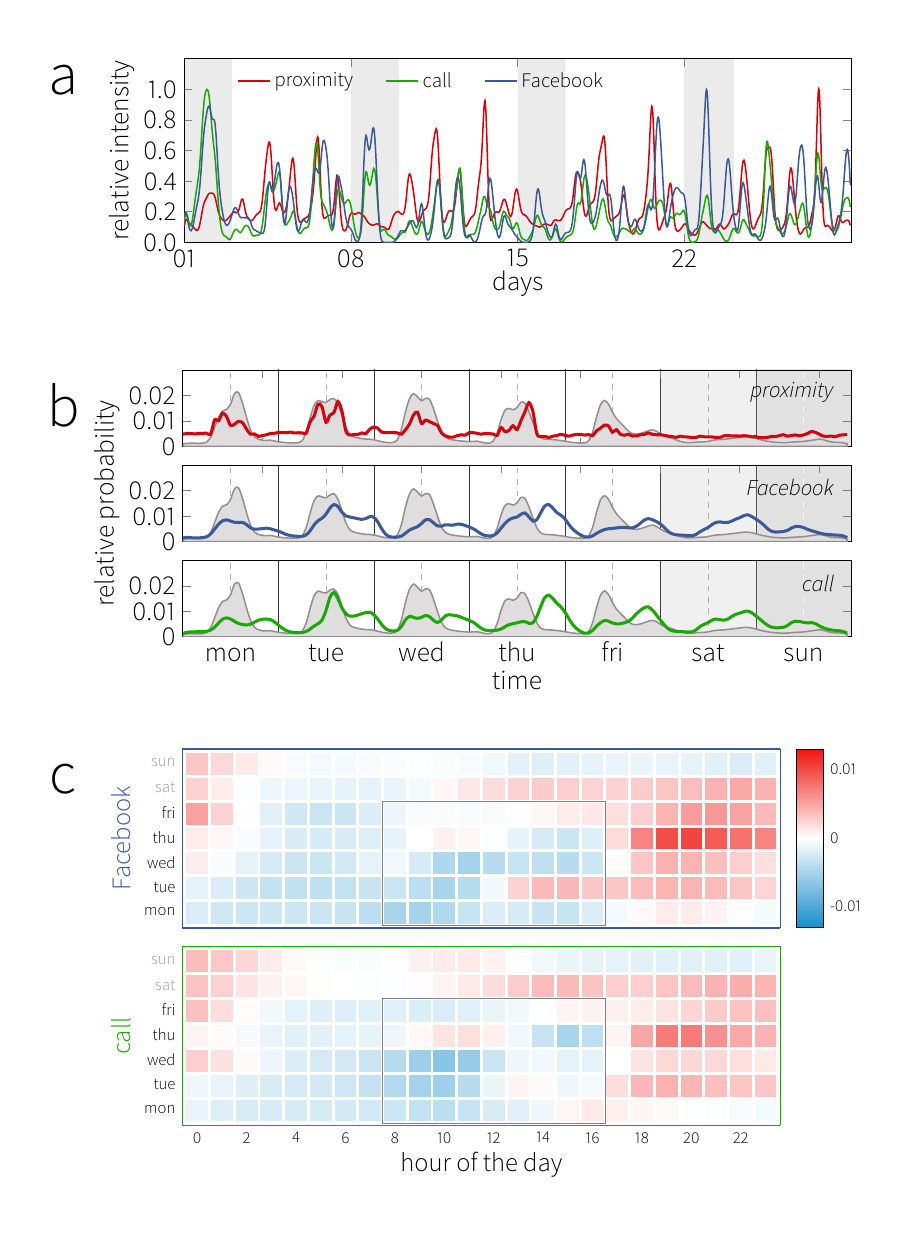}\\
\caption{\textbf{Contact activity of central individuals in the proximity network.}
a) Relative probability that a central member based on a specific channel has physical contact with any other individual.
b) Relative probability of physical contacts, compared to the distribution of the population average (grey curves).
Dark regions indicate weekends.
c) Heat maps showing the difference in the contact activity between the online elite and the proximity driven for each hour in a week.
The periods for \emph{working hours} are surrounded by the grey frame.
In all cases, the 10 most central individuals are considered, and results are aggregated over a four month period between February and May 2014 (inclusive).
\label{fig:contact_pattern}}
\end{figure}

On the contrary, phone calls and Facebook activity networks show the well-known rich structure of social networks, and therefore we are allowed to apply higher order centrality measures.
Results here are based on closeness centrality, which measures the average geodesic distance of an individual from the rest of the network.
More precisely, we rank students according to their closeness in the communication networks and select those with the highest value, but it should be noted that we obtain the same qualitative results when the selection is based on degree or $k$-coreness, a metric which has proved to be important indicators of influencers and spreaders in social networks\cite{kitsak2010identification,salathe2010high}.
Central individuals within each different network show distinct activity patterns with respect to their physical contacts.
Figure~\ref{fig:contact_pattern}a illustrates the relative intensity of physical contacts for the 10 most central individuals in the different networks over a period of a month (February 2014).
Although the curves are affected by local events, some general trends can be seen.
The first observation is that participants that are central in the intimate network are active and interact with the population primarily during the day (morning, noon and early afternoon) and show more regular interaction patterns (following the circadian rhythm), confirming that this channel has many interactions driven by daily schedules.
On the contrary, individuals central in the communication networks display increased physical activity during the weekends, evenings and nights, and show much less periodicity.
Motivated by these observations, we will refer to the central individuals selected by the communication networks as the \emph{online elite} and to the rest of the network as \emph{proximity driven}.
The latter labeling is supported by the notion that central individuals in proximity networks are those who tend to follow similar daily and weekly patterns in their interactions as the rest of the population.

The average weekly activity pattern obtained from the raw contact list of four months (February to May), show a profound difference whe compared to the population average, as seen in Fig.~\ref{fig:contact_pattern}b.
Central members of the proximity network (i.e., the proximity driven) relative to the online elite engage actively with the population as a whole according to the circadian rhythm and weekly schedules: most contacts take place during the day while students attend classes, with decreased intensity in the night.
Furthermore, the activity pattern of these individuals is not only consistent with the population average, but they also display a periodic intensity, limited to weekdays.
However, the online elite shows high contact activity during the afternoon, night and during the weekend, irrespective of the communication channel they are selected by.

Figure~\ref{fig:contact_pattern}c depicts a more detailed comparison of the activity of the online elite and the proximity driven, illustrated by the difference in the relative frequency that an online elite or proximity driven member interacts with any other individual.
In the plots, each tile shows the relative frequency of physical interactions by the top ten members of the online elite during a specific hour of the week, minus the relative frequency of interactions including the proximity driven in the same hour.
We refer to the outlined hours in the working days as \emph{working hours}, to distinguish that period from the rest of the week, that is, from hours where most of the voluntary and social activities are expected to take place (\emph{social hours}).
The online elite shows decreased activity during working hours compared to the proximity driven, and they are more active in the evening and nights as well as during the weekend (especially in the period that corresponds to Saturday night).
Also note that for most of the days (from Monday to Saturday), hours in the early morning do not display significant differences from the proximity driven, suggesting that the behavior of the online elite deviates from the rest of the population predominantly in working hours and nights.

% Figure 3

\subsection*{Regularity}
\begin{figure}
\centering
\includegraphics[width=0.5\textwidth]{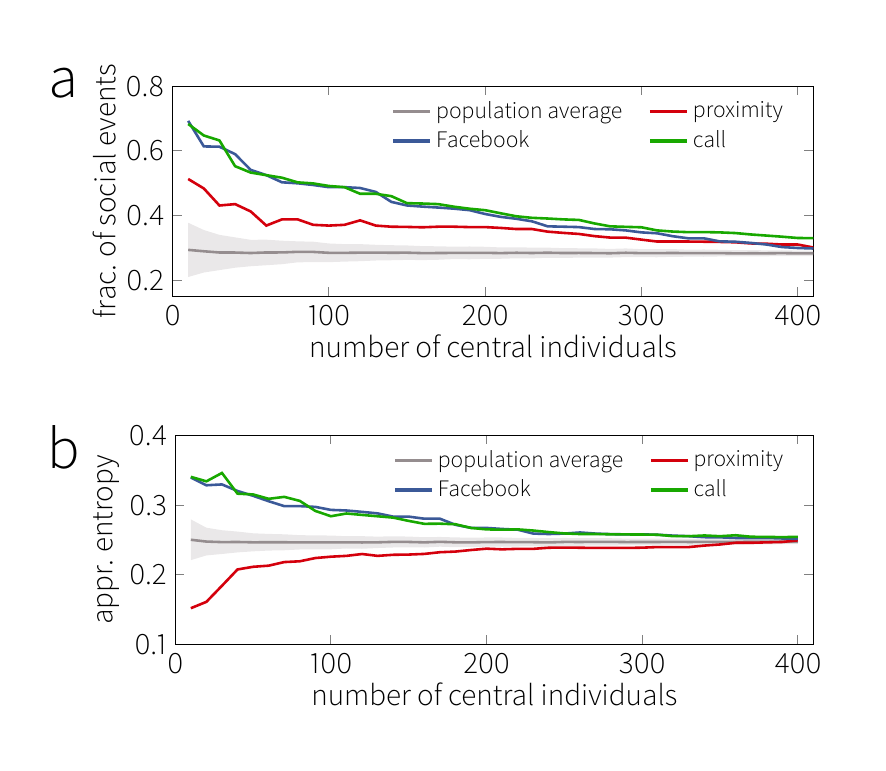}\\
\caption{\textbf{Activity during social hours and regularity of the online elite and the proximity driven.}
a) Median number of physical contact during social hours.
b) Approximate entropy of the contact activity.
Grey line denotes the population median with error bands representing lower and upper quartiles.
All data is calculated over a four month period between February and May 2014.
\label{fig:regularity}}
\end{figure}

The interaction frequency patterns in Fig.~\ref{fig:contact_pattern} indicate distinct behaviors in terms of activity as well as regularity in the case of the online elite and the proximity driven, respectively.
AWe quantify active periods and regularity in Fig.~\ref{fig:regularity} for groups consisting of a varying number of central individuals.
In case of the population average, curves represent the average of median values over a sample of 1\,000 randomly chosen groups.
First, we compare the fraction of contact events that take place during social hours to the population average for all three channels.
The online elite is characterized by a high fraction of contact events during social hours, and the difference does not vanish even for a group of 300 individuals, that is, almost 40\% of the population (Fig.~\ref{fig:regularity}a).
Proximity central individuals are also more active during social hours, however, they show less deviation from the population average.
Note that although the period of social hours is longer than that of the working hours, and therefore contacts have comparably higher propability to fall in social hours than to working hours, we merely focus on the relative behavior of the online elite and the proximity driven.

To measure regularity of the activity patterns, we calculate the approximate entropy of the relative frequency of contact events through a four month period.
Approximate entropy (ApEn) quantifies the level of irregularity in time series, comparing it to a completely periodic signal\cite{pincus1991approximate,pincus1991regularity}.
We chose ApEn due to its robustness against noise and because it can be efficiently computed from limited data.
Results are shown in Fig.~\ref{fig:regularity}b with sampling length of $m = 2$ and filter level of $r = 0.25$, however, results are robust with respect to the choice of $m$ or $r$.
Here we observe a strong effect: proximity based central individuals have an ApEn value that is even below the population average meaning that these individuals are more regular than the average.
On the other hand, the online elite shows sign of high irregularity for a large range of group sizes, starting with an ApEn that is 25\% higher than the population average.
The difference in the regularity measure of the online elite and the proximity driven vanishes only above the size of 200 individuals (approximately 40\% of the population).

% Figure 4

% discussion
\section*{Discussion}

With advances in technology, humans have started to use a variety of channels for communication, in addition to the physical presence in the proximity of others.
It is known that the different networks of interactions (physical contacts, online social media, phone calls, etc) correspond to different types of communication and can be the proxy for the strength of the social ties.
Due to the varying function of these communication modes, one might expect that the different channels favor different behavioral patterns in order for individuals to achieve a key structural position.
Here we made a contribution to the understanding of differences among individuals that are central in two fundamentally diverse environments: the network of physical proximity contacts and in digital communication networks (Facebook interactions and phone calls).
By locating central members within all three networks in the same coherent population, we find that the central members are described by qualitatively different presence and activity patterns in the physical contact networks.

The most central members of the population with respect to physical contacts, interact with others in a regular manner: they are most active during official schedule of a week (working hours) and follow a rather periodic activity pattern.
Therefore, their interactions can be easily predicted as they are limited by circadian rhythm and weekly schedules.
On the contrary, those central in the communication network (the online elite), display increased activity during periods of time outside working hours, that is, during social events not restricted by schedules.
The online elite also shows more irregular interaction activity and are therefore more difficult to predict.
In case of an infectious disease, predictive or preventive strategies that build upon the observation and recording of prevalence and infections, the fundamentally different behavior of the online elite and the proximity central should be taken into consideration.

While communication networks show low level of similarity both in strength of ties as well as structurally, the idiosyncratic behavioral patterns of the online elite illustrates how these channels can be applied to understand surprising aspects of our social interactions and to infer behavioral differences regarding real-world physical contacts.

\begin{methods}
\section*{Methods}

\subsection*{Data}
Data was collected during the Copenhagen Network Study (CNS) between 2012 and 2014\cite{stopczynski2014measuring}, and the results presented in this paper are obtained by analyzing the period from February to May 2014.
During the experiment, various data was collected from 1\,000 smartphones handed out to students of the Danish Technical University.

Due to the nature of the data and our methodological choices, these results are subject to various limitations that we discuss in the following.
First, the is a fraction of students with missing data resulting in low data quality.
To avoid working with structurally biased networks due to data loss, we selected a subset of students based on their coverage of proximity data: during the period of February - May 2014, we considered participants with signals in at least 60\% of the total time.
After the above filtering of the data, the size of the population considered in this paper is 532.

\subsection*{Networks}
From the CNS data, we built three types of networks: physical proximity networks are based on the Bluetooth scans of the devices.
These networks can be thresholded by the received signal strength index (RSSI) to obtain proximity networks with a distance of 1m (by setting \mbox{RSSI $>$ -75 dBm}).
Facebook feed and phone calls are used to create the communication networks: all interactions inside the population of 532 individuals are aggregated and static weighted networks are constructed.

\subsection*{Central groups}
In each network, we select central individuals, i.e., central groups of size $n$, by ranking the participants by a centrality measure and considering the $n$ ones with the lowest rank.
In case of proximity network, students are ranked by the total time spent in the proximity of others, while target groups in the communication networks are selected by their closeness centrality.
For participant $i$, the closeness centrality is defined by:
\[
C_C(i)= \frac{N - 1}{\sum_{j\ne i} d_{ij}},
\]
where $N$ is the number of participants and $d_{ij}$ denotes the geodesic distance between participant $i$ and $j$, i.e., the lowest number of steps to reach $j$ from $i$.
In case of a disconnected graph, $d_{ij}$ is defined to be $N$.

\end{methods}

\bibliographystyle{naturemag}
%\bibliography{bibliography}

%% Here is the endmatter stuff: Supplementary Info, etc.
%% Use \item's to separate, default label is "Acknowledgements"

\begin{addendum}
 \item This work was supported a Young Investigator Grant from the Villum Foundation (High Resolution Networks, awarded to S.L.), and interdisciplinary UCPH 2016 grant (Social Fabric).
 Due to privacy implications we cannot share data but researchers are welcome to visit and work under our supervision.
 \item[Competing Interests] The authors declare that they have no competing financial interests.
 \item[Correspondence] Correspondence and requests for materials should be addressed to EM~(email: enmo@dtu.dk).
\end{addendum}

%%
%% TABLES
%%
%% If there are any tables, put them here.
%%

\end{document}